\documentclass[a4paper,fleqn,usenatbib]{mnras}

\usepackage[T1]{fontenc}
\usepackage{ae,aecompl}
\usepackage{xcolor}

\usepackage{graphicx}	
\usepackage{amssymb}	
\usepackage{pdflscape}	
\usepackage[english]{babel}

\newcommand{\msol}{\mathrm{M}_{\rm \odot}}

\newcommand{\methylcyanide}{CH$_3$CN}

\newcommand{\methanol}{CH$_3$OH}

\title[Self-gravitating disc models for massive stars]{Self-gravitating disc candidates around massive young stars}

\author[D. H. Forgan et al.]
{D.~H.~Forgan$^1$\thanks{Contact e-mail: \href{mailto:dhf3@st-andrews.ac.uk}{dhf3@st-andrews.ac.uk}},
J.~D.~Ilee$^2$, 
C.~J.~Cyganowski$^1$,
C.~L.~Brogan$^3$ and
T.~R.~Hunter$^3$
\vspace{0.2cm} \\
$^{1}$SUPA, School of Physics \& Astronomy, University of St Andrews, North Haugh, St Andrews, Scotland, KY16 9SS, UK\\
$^{2}$Institute of Astronomy, Madingley Road, Cambridge CB3 0HA, UK\\
$^{3}$NRAO, 520 Edgemont Rd, Charlottesville, VA 22903}

\date{Accepted 2016 July 29. Received 2016 July 29; in original form 2016 May 27}

\pubyear{2016}

\begin{document}
\label{firstpage}
\pagerange{\pageref{firstpage}--\pageref{lastpage}}
\maketitle

\begin{abstract}
There have been several recent detections of candidate Keplerian discs around massive young protostars.  Given the relatively large disc-to-star mass ratios in these systems, and their young ages, it is worth investigating their propensity to becoming self-gravitating.  To this end, we compute self-consistent, semi-analytic models of putative self-gravitating discs for five candidate disc systems.  Our aim is not to fit exactly the observations, but to demonstrate that the expected dust continuum emission from marginally unstable self-gravitating discs can be quite weak, due to high optical depth at the midplane even at millimetre wavelengths. In the best cases, the models produce ``observable'' disc masses within a factor of <1.5 of those observed, with midplane dust temperatures comparable to measured temperatures from molecular line emission.  We find in two cases that a self-gravitating disc model compares well with observations.  If these discs are self-gravitating, they satisfy the conditions for disc fragmentation in their outer regions.  These systems may hence have as-yet-unresolved low mass stellar companions, and are thus promising targets for future high angular resolution observations.
\end{abstract}




\section{Introduction}
\label{sec:introduction}

Massive stars have a large impact on the evolution of the interstellar medium, not only on their immediate vicinity due to energetic stellar winds, ionising radiation and outflows, but also on galactic scales via the continual processing of heavy elements released when they die as supernovae.  Therefore, a good understanding of the formation of massive stars is crucial for many areas of astrophysics.  However, despite low mass star formation being relatively well understood \citep[e.g.][]{shu_1987}, there is little consensus on how exactly high mass ($M_{\star}  >  8$\,M$_{\odot}$) stars form.  

\smallskip

Observationally, difficulties arise from the fact that the pre-main-sequence phase of massive stars is very short ($\sim$10$^{4}$   --  a  few  $\times$ 10$^{5}$\,years, e.g.\ \citealt{davies_2011}) thus providing little opportunity to directly observe massive young stellar objects (MYSOs) due to high extinctions and the small number of candidate objects.  However, despite these difficulties, recent years have seen a steady rise in the number of detections of candidate Keplerian discs around massive young protostars \citep[see e.g.][]{hunter_2014, johnston_2015, zapata_2015, chen_2016, Ilee2016}.  

\smallskip

From a theoretical perspective, circumstellar discs provide a mechanism for accretion that overcomes the feedback from the central protostar, via channelling material along the equatorial plane \citep[see the models of, e.g.,][]{krumholz_2009, kuiper_2011, klassen_2016}.  A common feature of these models is the prediction of disc instability \citep{vaidya_2009} leading to significant non-axisymmetric structure, caused by high accretion rates through the disc, large disc-to-star mass ratios and self-gravity \citep[see, e.g.,][]{vorobyov_2010}.  However, such details have been below the spatial resolution limits of previous observational campaigns, due in part to the high angular resolution required to observe such structure, even for nearby low mass systems \citep[e.g.][]{douglas_2013, dipierro_2014}.

\smallskip

The inferred physical parameters of disc candidates around massive stars, in particular the estimated masses, suggest that they are in fact self-gravitating discs.  Simulations of low mass star formation suggest that during the very earliest epochs, protostars are likely to be surrounded by discs of comparable mass, and hence disc self-gravity is important.  In the first-core second-core paradigm of low mass star formation \citep{larson_1969,Masunaga_1}, it seems that for protostellar collapse in the presence of rotation, the first-core is in fact the disc, with the second-core forming the protostar \citep{Bate2010}.  If massive star formation proceeds in a similar format to low mass star formation, albeit in a scaled-up manner, then we should expect similar behaviour.

\smallskip

Self-gravitating discs can produce highly effective angular momentum transport through low mode spiral density waves at very high masses \citep{Laughlin1996} and through gravito-turbulence, which produces a relatively strong turbulent viscosity at a range of disc masses \citep{gammie_2001}.  The self-gravitating phase is likely to be relatively brief, but provides an attractive means for loading mass onto the central massive star, and may have significant impact on the subsequent dynamical and chemical evolution of the disc \citep[e.g.][]{ilee_2011, evans_2015}.  As self-gravitating discs can be described in a pseudo-viscous manner under certain conditions \citep{forgan_2011}, it is relatively simple to compute quasi-steady disc models given the observed accretion rate, stellar mass and the disc inner and outer radii.  

\smallskip

In this paper, we compute such models for five massive stars with reported observations of relatively massive circumstellar disc-like structures.  We show that in several cases, the observational constraints on disc properties strongly suggest that self-gravity is playing a key role in mediating angular momentum transport in these discs.  Further to this, some of the discs display accretion rates sufficiently high to prevent quasi-steady self-gravitating disc solutions.  Our models therefore predict these systems could be undergoing disc fragmentation into bound objects, with expected fragment masses above the hydrogen burning limit, i.e.\ low mass stars.  The paper is structured as follows: we outline the self-gravitating disc model in Section \ref{sec:method}.  We apply this model to the observations on an object-by-object basis in Section \ref{sec:results}.  Finally, our conclusions are presented in Section \ref{sec:conclusions}.

\begin{table*}
	\begin{minipage}{0.96\textwidth}
	\centering
	\caption{Star and disc parameters adopted in the models.}
	\label{tab:properties}
	\begin{tabular}{lccccc} 
	\hline
         Object     & Stellar Mass & Stellar Radius & Effective Temperature     & Disc Inner Radius     &  Disc Outer Radius     \\
         
                    & $M_{\star}$ (M$_{\odot}$)   & $R_{\star}$ (R$_{\odot}$) & $T_{\mathrm{eff}}$ (K)        &  $R_{\mathrm{in}}$ (au)    & $R_{\mathrm{out}}$ (au)     \\ 
    \hline
        G11.92$-$0.61 MM1$^{a}$       &  34   & 9   & 6300    &  21   &  1200     \\
        NGC 6334 I(N) SMA1b$^{b}$     &  20   & 15  & 6100    &  22   &  800      \\
        AFGL 4176 mm1$^{c}$               &  25   & 10  & 6200    &  31   &  2000     \\
        IRAS 16547$-$4247$^{d}$       &  20   & 15  & 6100    &  20   &  800      \\
        IRAS 20126$+$4104$^{e}$       &  12   & 35  & 6000    &  20   &  1200     \\
	\hline
	\end{tabular}
	\begin{flushleft}
	\small{$a$: \citet{Ilee2016}, $b$: \citet{hunter_2014}, $c$: \citet{johnston_2015}, $d$: \citet{zapata_2015}, $e$: \citet{chen_2016}.}\\
	\end{flushleft}
\end{minipage}
\end{table*}

\section{Method}
\label{sec:method}

\subsection{Quasi-steady Self-Gravitating Disc Models}

Discs are gravitationally unstable if the Toomre Parameter \citep{toomre_1964}:
\begin{equation}
    Q = \frac{c_s \kappa_{\rm epi}}{\pi G \Sigma} \sim 1, \label{eq:Q}
\end{equation}
where $c_s$ is the local sound speed, $\Sigma$ is the local surface density and $\kappa_{epi}$ is the epicyclic frequency (in Keplerian discs, this is equal to the angular frequency $\Omega$). This criterion is sufficient to produce axisymmetric instabilities in the disc.  Numerical simulations indicate that non-axisymmetric instabilities proceed when $Q\sim 1.5-1.7$ \citep[see e.g.][]{durisen_2007,kratter_2016}.

\smallskip

We can construct a simple one dimensional self-gravitating disc model assuming that the angular momentum transport is pseudo-viscous \citep{clarke_2009,rice_2009}.  We use the $\alpha$-prescription \citep{shakura_1973}:
\begin{equation}
    \nu = \alpha c_s H,
\end{equation}
where $\nu$ is the turbulent viscosity, and $H$ is the disc scale height:
\begin{equation}
    H = \frac{c_s}{\Omega} \approx \frac{c^2_s}{\pi G \Sigma}.
\end{equation}

We are in the limit that the non-self-gravitating and self-gravitating expressions for the scale height are approximately equal thanks to Equation (\ref{eq:Q}). This pseudo-viscous approach is acceptable if the angular momentum transport is locally determined, and the disc aspect ratio $H/r$ remains low (of order 0.1), or equivalently the disc-to-star mass ratio is below 0.5 \citep{forgan_2011}.  We shall see that the disc masses derived for the five massive stars we study are sufficiently low to justify assuming local angular momentum transport.

\smallskip

We construct our disc model in the same manner as \citet{forgan_2011a,forgan_2013,forgan_2013a}.  We assume a fixed $Q=2$, and that the accretion rate, $\dot{M}$, is constant across all radii:

\begin{equation}
    \dot{M} = 3 \pi \nu \Sigma = \frac{3 \pi \alpha c^2_s \Sigma}{\Omega}.
\end{equation}
The value of $\alpha$ at each radius is determined assuming local thermodynamic equilibrium:
\begin{equation}
    \alpha = \frac{4}{9\gamma(\gamma-1) \beta_c},
\end{equation}
where we assume $\gamma=5/3$.  We obtain the dimensionless cooling parameter
\begin{equation}
    \beta_c = t_{\mathrm{cool}} \Omega^{-1}
\end{equation}
by assuming the disc cools according to
\begin{equation}
    \dot{u} = -\frac{u}{t_{\mathrm{cool}}} = \frac{\sigma_{\mathrm{SB}} (T^4-T^4_{\mathrm{irr}})}{\tau + \tau^{-1}},
\end{equation}
where $u$ is the local internal energy per unit mass, $\sigma_{\mathrm{SB}}$ is the Stefan-Boltzmann constant, $\tau = \Sigma \kappa$ is the local optical depth (from a grey opacity $\kappa$), $T$ is the disc temperature and $T_{\mathrm{irr}}$ is the temperature of the local environment due to irradiation from the central star.  We calculate $\Omega$ from the standard Keplerian expression:
\begin{equation}
    \Omega(r) = \sqrt{\frac{G M_{\mathrm{enc}}(r)}{r^3}},
\end{equation}
where we account for the disc self-gravity by replacing $M_{\star}$ with the total mass within a radius $r$, $M_{\mathrm{enc}}(r)$.  The system of equations is hence closed, and we solve via iteration of the surface density at each radius. Therefore, for a given value of the star mass $M_{\star}$ and accretion rate $\dot{M}$, we deliver a self-consistent self-gravitating disc model for a given inner and outer radius, and we can obtain the disc mass by integrating the surface density. 

\subsection{Stellar Irradiation}

The local irradiation temperature is dominated by the massive central protostar.  We assume a standard Stefan-Boltzmann relationship between equilibrium temperature and distance from the star \citep[cf][]{hayashi_1981}:
\begin{equation}
    T_{\mathrm{irr}}(r) = 280 \left(\frac{M}{\msol}\right) \left(\frac{r}{1 \,\mathrm{au}}\right)^{-1/2}.
\end{equation}

The above assumes a standard main sequence mass luminosity relation.  However, such a simple prescription vastly overestimates the stellar irradiation from massive young protostars.  \citet{hosokawa_2009} determine that at the large accretion rates measured in MYSOs, the entropy generated at the accretion shock cannot be radiated away efficiently due to long cooling times, resulting in a swelling of the stellar radius and a corresponding reduction in the effective temperature $T_{\mathrm{eff}}$.  As such, we determine stellar radii and effective temperatures for each object from the models of \citet{hosokawa_2009}, based on the stellar mass, and rescale the above equation to ensure that $T_{\mathrm{irr}}(r=R_{\star})=T_{\mathrm{eff}}$.  The star and disc properties assumed for each object are listed in Table \ref{tab:properties}, however we note from test simulations that our results are only sensitive to these parameters at low disc radii. At large distances from the star where fragmentation can occur, we find little effect on the resulting disc structure.

\subsection{Fragmentation and Fragment Masses}

We test for the propensity of the disc to fragment into bound objects using the Jeans mass formalism of \citet{forgan_2011a,forgan_2013}.  The local Jeans mass inside a spiral density wave in a self-gravitating disc is given by:
\begin{equation} M_J = \frac{4\sqrt{2}\pi^3 }{3G}\frac{Q^{1/2} c^2_s H}{\left(1 + \frac{\Delta \Sigma}{\Sigma}\right)}, \label{eq:mjeans}
\end{equation}
where we use the empirical relation from \citet{rice_2011} (see also \citealt{Cossins2008}):
\begin{equation}
     \left<\frac{\Delta \Sigma_{\mathrm{rms}}}{\Sigma}\right> = 4.47 \sqrt{\alpha}. 
\end{equation}

We compute the time derivative $\dot{M}_J$ and identify regions where this quantity is large and negative.  In these regions, surface density perturbations produced by the gravitational instability are most likely to result in a parcel of disc material that is Jeans unstable and will begin collapsing into a disc fragment.

\subsection{Observed Dust-derived Mass}

We can compare the disc mass obtained from the model to observations by computing the predicted dust continuum emission.  We estimate the observed flux density at frequency $\nu$ using (cf\ \citealt{forgan_2013a}):
\begin{equation}
F_{\nu}(r) dr= \left\{
\begin{array}{l l }
\frac{2k}{c^2D^2} \nu^2 \kappa(\nu) \Sigma(r) T(r) 2\pi r dr & \tau \leq 1  \\
\frac{2k}{c^2D^2} \nu^2  \frac{T(r)}{\tau^{1/4}} 2\pi r dr & \tau > 1,
\end{array} \right.
\label{eq:modflux}
\end{equation}
where the optical depth $\tau = \Sigma \kappa$ and $D$ is the system distance.  This approach reflects the fact that self-gravitating discs can become optically thick even at quite long wavelengths, and as such only the dust mass above the photosphere (with temperature $T_{phot}=\frac{T(r)}{\tau^{1/4}}$) is measurable \citep{greaves_2010}.  

\smallskip

The greatest source of uncertainty in these calculations is the opacity.  Where possible, we adopt the opacity used by the observers (assuming a gas-to-dust mass ratio $R_{g}=100$ where necessary). We then convert this into an observed disc mass using
\begin{equation}
M_{\mathrm{disc}} = \frac{D^2 F_\nu c^2}{2 \kappa(\nu) k \nu^2 T_{\mathrm{dust}}}, \label{eq:Mdisc_dust}
\end{equation}
where $\kappa(\nu)$ is a frequency-dependent opacity law, and $T_{dust}$ is the inferred dust temperature.  We adopt the same opacity law and dust temperature as used for each observation, and we also adopt a common observer assumption that the disc is optically thin, which is unlikely to be the case for massive self-gravitating discs.  Note also that assuming a single dust opacity, a common practice when recovering disc masses, is also problematic when it is evident that opacities will be radially dependent.  Equations (\ref{eq:modflux}) and (\ref{eq:Mdisc_dust}) assume the frequency is sufficiently short (or wavelength sufficiently long) that the radiation is emitted from the Rayleigh-Jeans tail of the blackbody spectrum.  Some observers do instead use the full Planck function in equation (\ref{eq:Mdisc_dust}), but this has a negligible effect on $M_{\mathrm disc}$.

\section{Results}
\label{sec:results}

\subsection{G11.92$-$0.61 MM1}

\begin{figure*}
  $\begin{array}{cc}
  \includegraphics[width=\columnwidth]{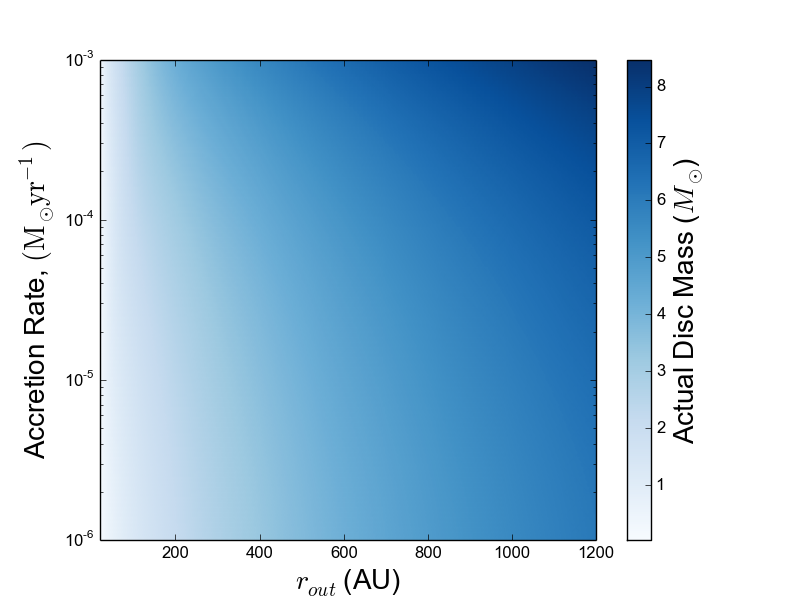} &
  \includegraphics[width=\columnwidth]{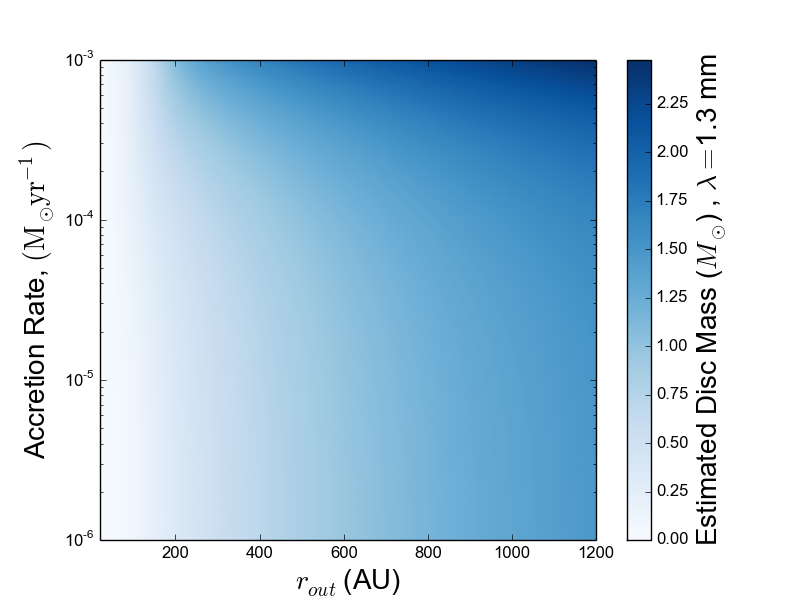} \\
  \includegraphics[width=\columnwidth]{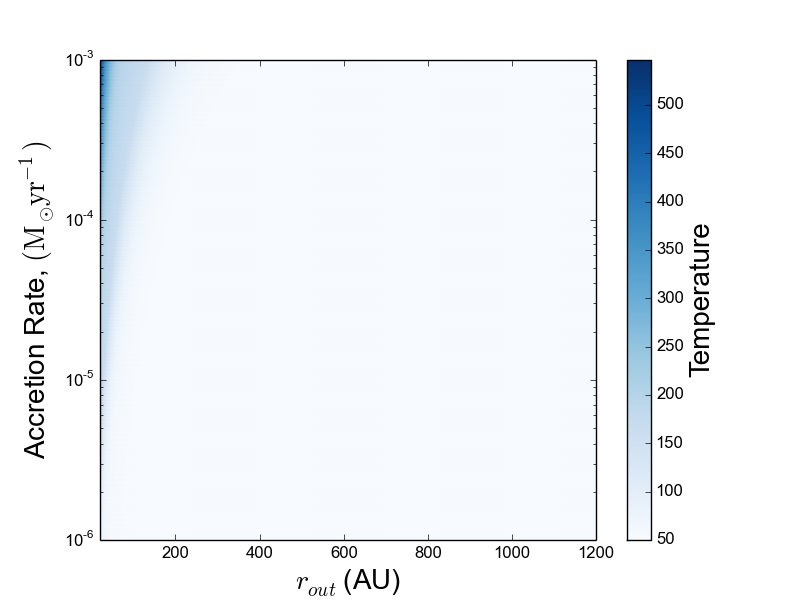} &
  \includegraphics[width=\columnwidth]{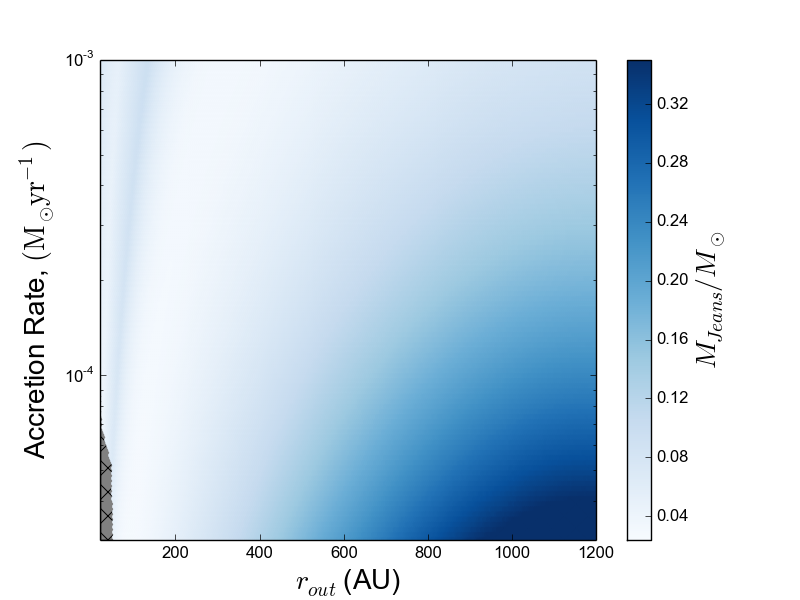} \\
  \end{array}$
  \caption{Self-gravitating disc models for MM1, assuming a central object mass of $34\,\msol$.  The inner disc radius is fixed at 21\,au. Each accretion rate and outer radius point represents a self-gravitating disc model, with colours representing the value of a given variable for that model (\textit{top left} - disc mass interior to $r_{\rm out}$; \textit{top right} - disc mass measured from dust continuum flux; \textit{bottom left} - midplane temperature; \textit{bottom right} - local Jeans mass inside a spiral density wave, i.e.\ the local mass of fragments (where fragmentation occurs)). Hashed regions in the lower right plot indicate disc models that do not produce fragments.}
  \label{fig:MM1_34msol}
\end{figure*}

\noindent This system (hereafter MM1) has been shown by \citet{Ilee2016} to fit a Keplerian rotation profile, with an enclosed mass of approximately 30--60\,$\msol$ within $\sim 1200$\,au (at a distance of 3.37\,kpc).  The disc mass is around $2-3\,\msol$, estimated from the continuum dust emission. The dust is assumed to be optically thin but with an opacity correction based on the observed millimetre continuum brightness temperature, which has an effect of order 10\% on the mass estimate.  MM1 drives a bipolar molecular outflow with a relatively short dynamical time of $\leq$ 10,000 years \citep{cyganowski_2011sma}, suggesting this is a particularly young system.  Along with its high disc to star mass ratio of $\gtrsim 0.035$, this suggests that MM1 could be a self-gravitating disc system.

\smallskip

We display outputs from the self-gravitating disc models for MM1 in Figure \ref{fig:MM1_34msol} assuming a central star mass of $34\,\msol$. The inner disc radius is set to 21\,au, i.e. at the boundary of the source's hyper-compact H\,{\sc ii} region, and we neglect the ionised material within.  

\smallskip

The top left panel of Figure \ref{fig:MM1_34msol} shows that for the disc parameters derived from this model, the true disc mass is between 5 -- 8\,$\msol$ depending on the accretion rate.  By assuming the dust temperature is 150\,K, we derive an ``observed'' disc mass (from equation \ref{eq:Mdisc_dust}) of around 1.5 -- 2\,$\msol$ (right panel of Figure \ref{fig:MM1_34msol}), which is close to that measured by \citet{Ilee2016} and \citet{cyganowski_2014}. This temperature is determined from the cool component of the \methylcyanide\ emission, and is reasonably consistent with midplane temperatures in the self-gravitating disc model, which are of order a few hundred Kelvin (bottom left panel of Figure \ref{fig:MM1_34msol}).  Our estimate excludes the mass of any surrounding envelope, which may explain why our model's ``observed'' mass is a slight underestimate.

\smallskip

If our model is correct, we predict the disc should also be in the process of fragmenting. The Jeans mass criterion is easily satisfied for accretion rates above around $3 \times 10^{-5} \, \msol \mathrm{yr}^{-1}$. The bottom right panel of Figure \ref{fig:MM1_34msol} shows the fragment mass in solar masses (for disc parameters where fragmentation occurs).  Generally, the Jeans mass exceeds the hydrogen burning limit of 0.08 $\msol$, and it is expected that fragments can accrete further mass from the surrounding disc shortly after birth. We would therefore expect the system to be forming low mass protostars in the outer regions of the disc, presumably containing their own subdisc systems (see e.g. \citealt{forgan_2016}). These objects are beyond the resolution of our observations at this time, but may be detectable with future high angular resolution campaigns \citep{cossins_2010,vorobyov_2013, Dong2016}. 

\smallskip

The formation of protostars is likely to launch jets.  It is tempting to speculate that the water maser emission observed in the outer regions of the MM1 disc \citep{moscadelli_2016} could be indicative of fragmentation, with the masers tracing the base of a disc fragment's jet column, but such masers are sufficiently common that their presence is at best circumstantial evidence.  

\subsection{NGC 6334 I(N)-SMA1b}

\begin{figure*}
  $\begin{array}{cc}
  \includegraphics[width=\columnwidth]{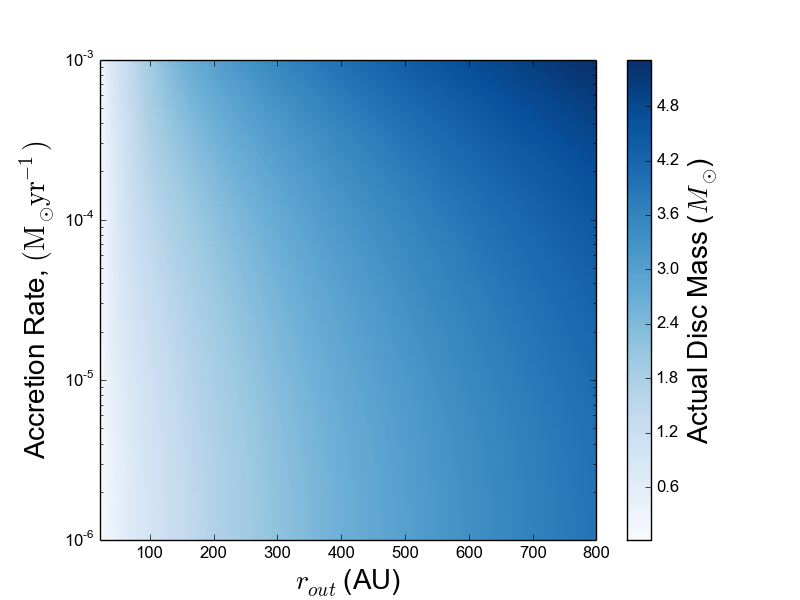} &
  \includegraphics[width=\columnwidth]{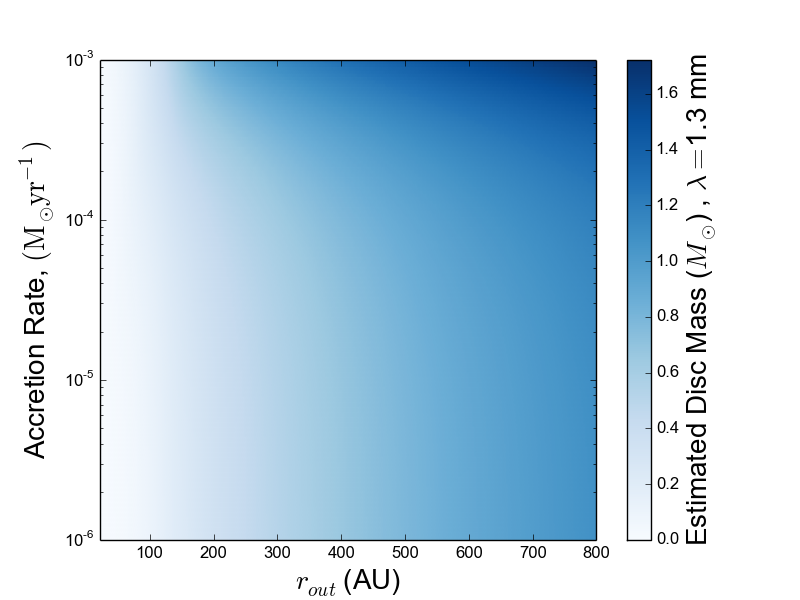} \\
  \includegraphics[width=\columnwidth]{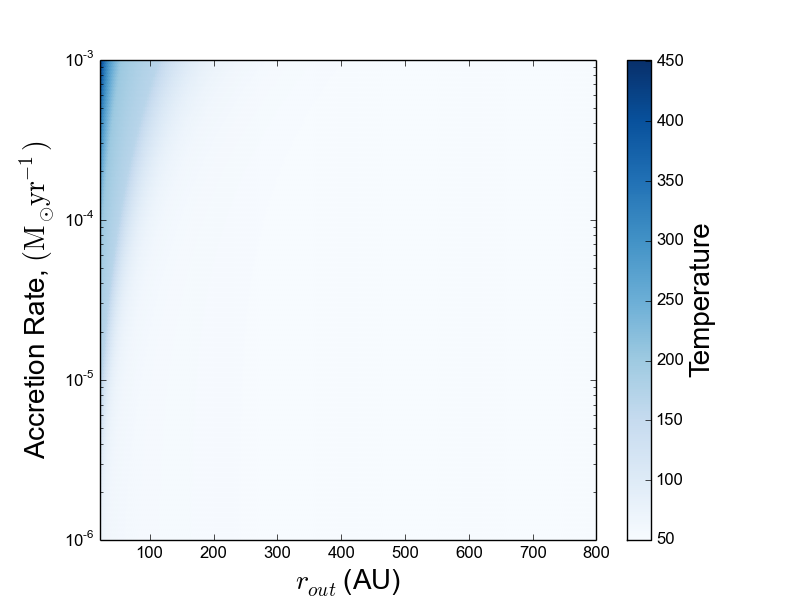} &
  \includegraphics[width=\columnwidth]{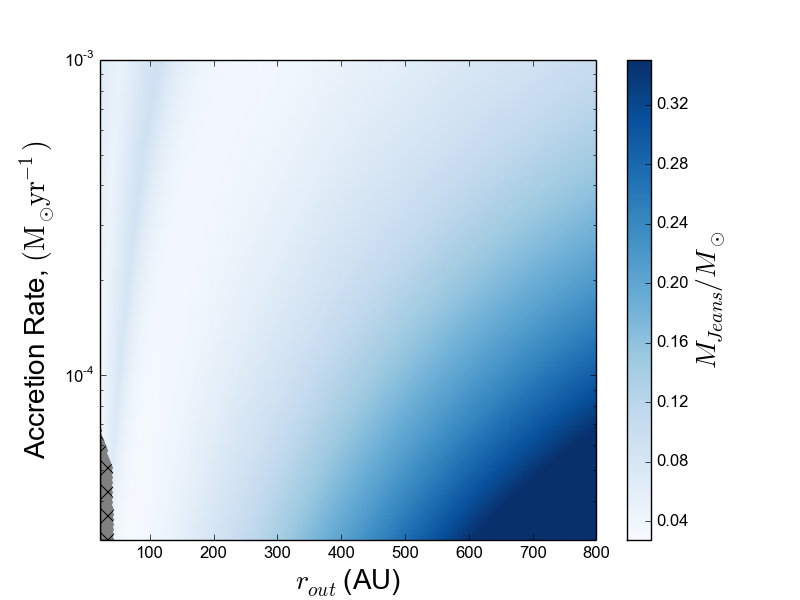} \\
  \end{array}$
  \caption{As Figure \ref{fig:MM1_34msol}, but for the SMA1b system.}
  \label{fig:INorth_20msol}
\end{figure*}

\noindent NGC 6334 I(N)-SMA1b (hereafter SMA1b) is a candidate disc system with enclosed mass of $10-30\,\msol$ within 800\,au (depending on the system inclination), at a distance of 1.3\,kpc \citep{hunter_2014}.  A velocity gradient traced by multiple species, perpendicular to an outflow traced by SiO(5-4) emission, is strong evidence for a disc configuration \citep{brogan_2009,hunter_2014}.

\smallskip


Figure \ref{fig:INorth_20msol} shows the resulting disc models for a 20\,$\msol$ star with an inner radius of 22\,au (based on the reported radius of the hyper-compact H\,{\sc ii} region).  The true disc masses derived in this case are approximately 4 $\msol$, although the mass derived from the dust continuum (assuming a dust temperature of 150\,K) is around 1.3--1.7\,$\msol$.  Given \citet{hunter_2014}'s estimate of $\sim $3--5\,$\msol$ from dust continuum emission (assuming optically thin emission, corrected for opacity according to the system's brightness temperature), our prediction lies just below their lower limit.  Again, extra mass in the surrounding envelope may explain the small discrepancy between our prediction and the observations, again suggesting that SMA1b may be a self-gravitating disc system, with the potential to form low mass stars via fragmentation.

\subsection{AFGL 4176-mm1}

\begin{figure*}
  $\begin{array}{cc}
  \includegraphics[width=\columnwidth]{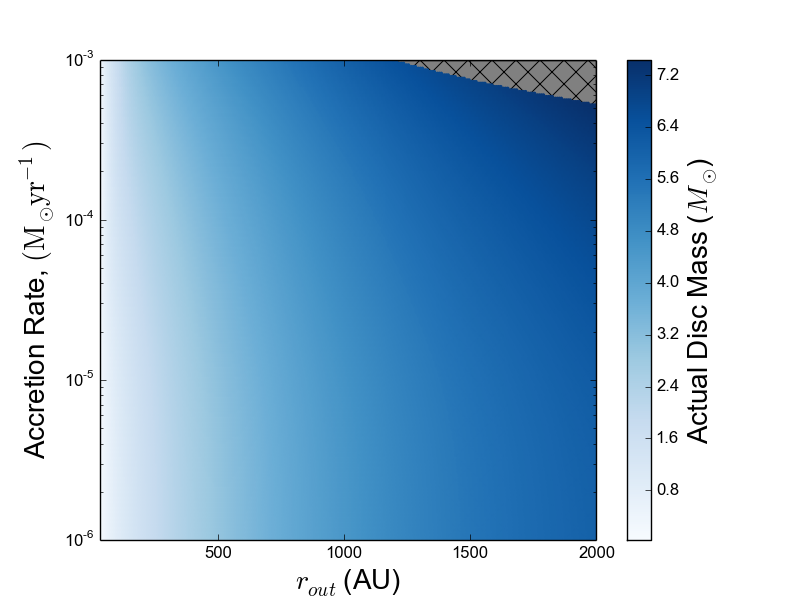} &
  \includegraphics[width=\columnwidth]{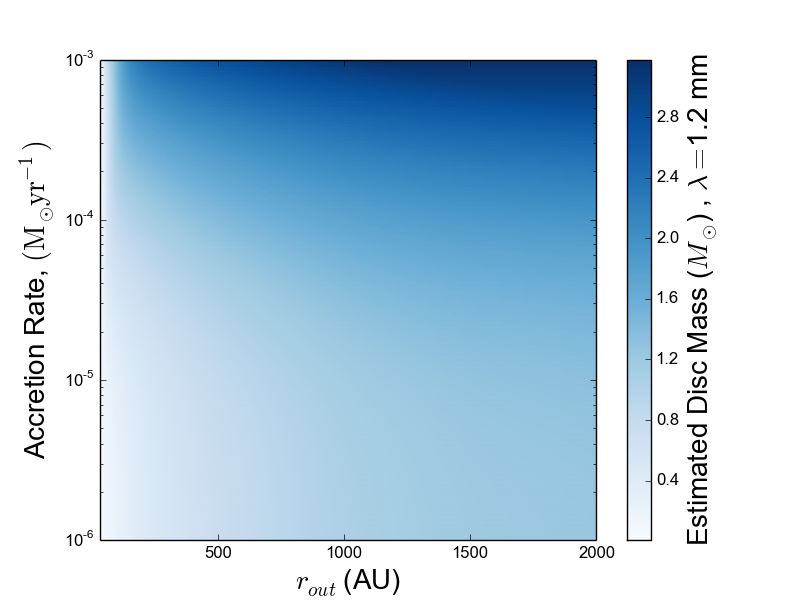} \\
  \includegraphics[width=\columnwidth]{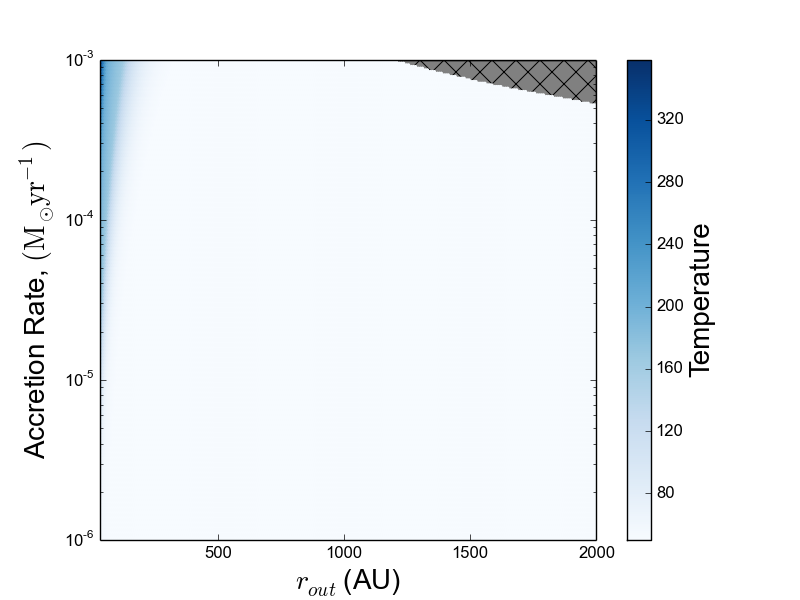} &
  \includegraphics[width=\columnwidth]{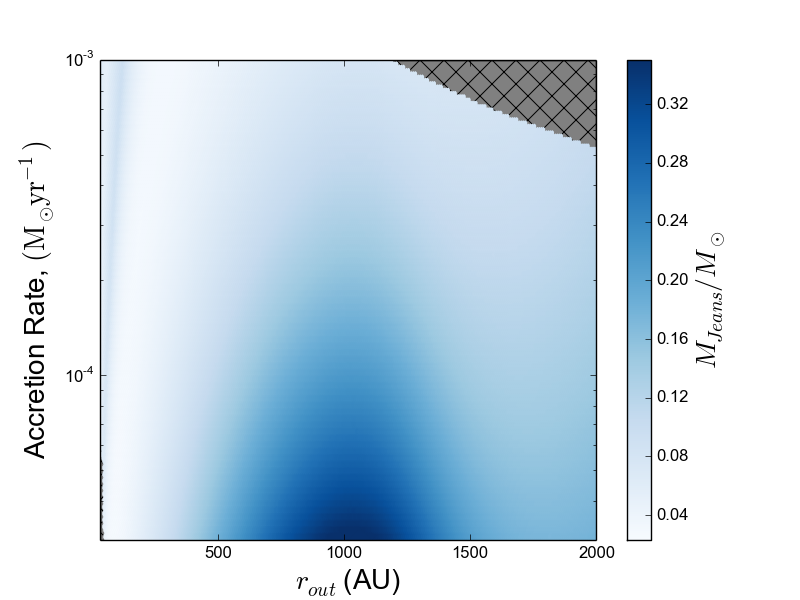} \\
  \end{array}$
  \caption{As Figure \ref{fig:MM1_34msol}, but for the AFGL 4176 mm1 system.  The grey hatched regions denote parameter spaces which do not produce viable self-gravitating disc solutions, due to irradiation preventing $Q$ from reaching the instability regime.  \citet{johnston_2015}'s derived accretion rate places AFGL4176 in the hashed region, hence we conclude that AFGL4176 is not gravitationally unstable.}
  \label{fig:AFGL4176}
\end{figure*}

\noindent This forming O-star has a Keplerian-like disc, with mass measured to be $\sim 8\,\msol$ assuming isothermal dust emission, and $12\,\msol$ when detailed radiative transfer modelling is employed \citep{johnston_2015}.

\smallskip

Our models (Figure \ref{fig:AFGL4176}), assuming a 25\,$\msol$ central star and an inner radius of 31 au, suggest that this system is probably not completely self-gravitating.  If the infall rate from the envelope modelled by \citet{johnston_2015} is correct, and the disc accretes at a similar rate ($\dot{M} =  4.6\times 10^{-4}\, \msol \mathrm{yr}^{-1}$), then even relatively weak irradiation prevents the Toomre $Q$ parameter from reaching low values, and as such the optically thin outer disc is stable against self-gravity.  This is shown by the grey hatched regions in Figure \ref{fig:AFGL4176}, which indicate that a consistent self-gravitating solution cannot be found.  Our true disc masses are slightly lower than \citet{johnston_2015}'s model fit to their data.   Our estimates of disc masses from optically thin continuum dust emission are lower still.  If we assume \citet{johnston_2015}'s adopted dust temperature of 190 K, and opacity law \citep{draine_2011}, we derive an observed mass of around 3\,$\msol$.

\smallskip

This discrepancy highlights the dangers of assuming dust emission is always optically thin. If we assume that the full continuum radiative transfer modelling carried out by \citet{johnston_2015} gives a faithful disc mass estimate, then we can compare their results to the true disc mass generated by the self-gravitating disc model, and we see that our true disc mass estimates are lower.  We can therefore conclude that the data for AFGL 4176 mm1 are consistent with a system that is sufficiently irradiated by external sources to prevent the gravitational instability activating, and has possibly just left the self-gravitating phase. 

\subsection{IRAS 16547$-$4247}

\citet{zapata_2015} presented observations of this system, indicating an enclosed mass of around 20--30\,$\msol$ within 1000\,au, and an upper limit on the disc mass of around 6\,$\msol$ from dust emission, assuming the dust is optically thin.

\begin{figure*}
  $\begin{array}{cc}
  \includegraphics[width=\columnwidth]{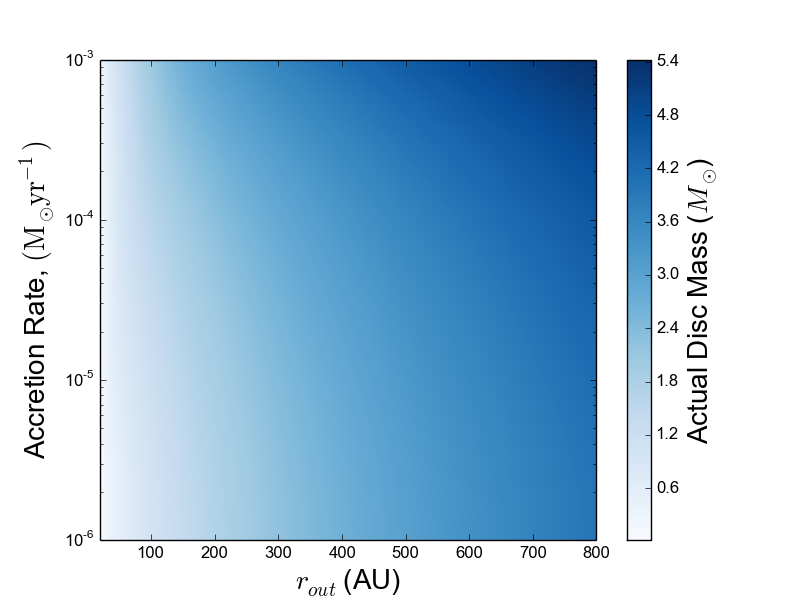} &
  \includegraphics[width=\columnwidth]{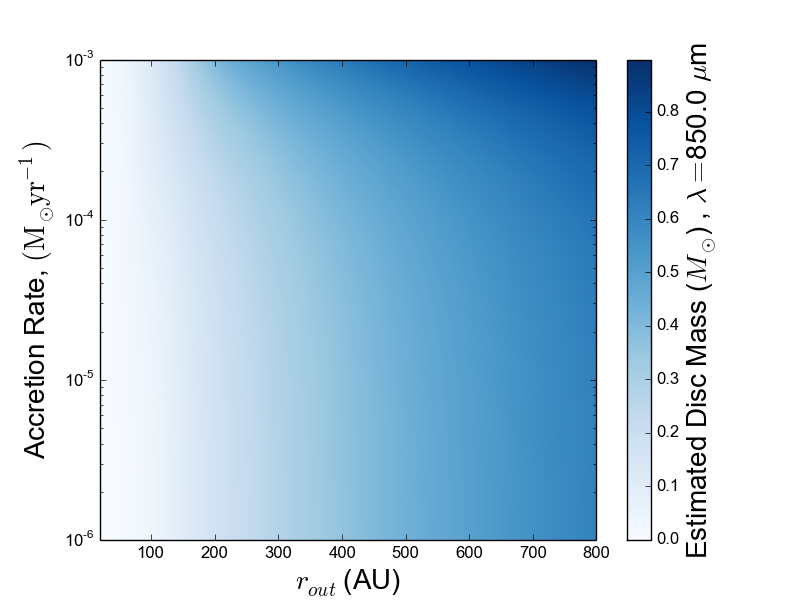} \\
  \includegraphics[width=\columnwidth]{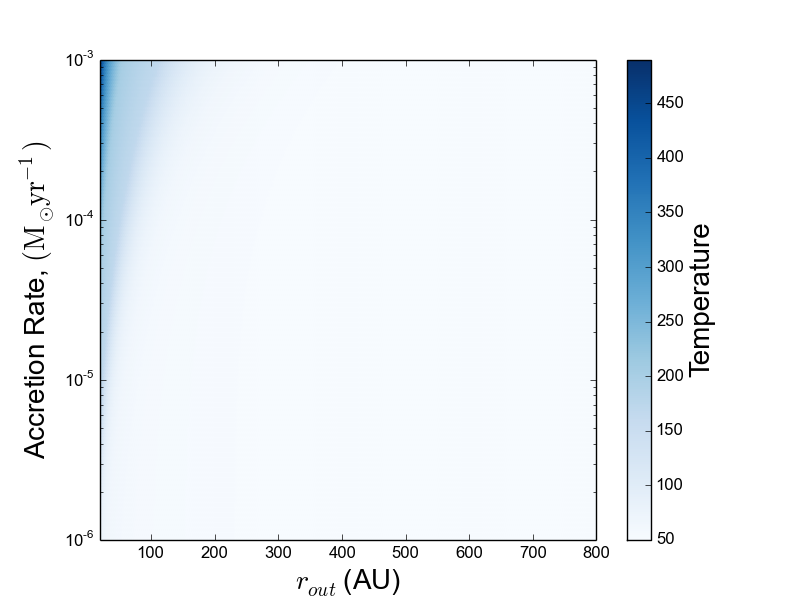} &
  \includegraphics[width=\columnwidth]{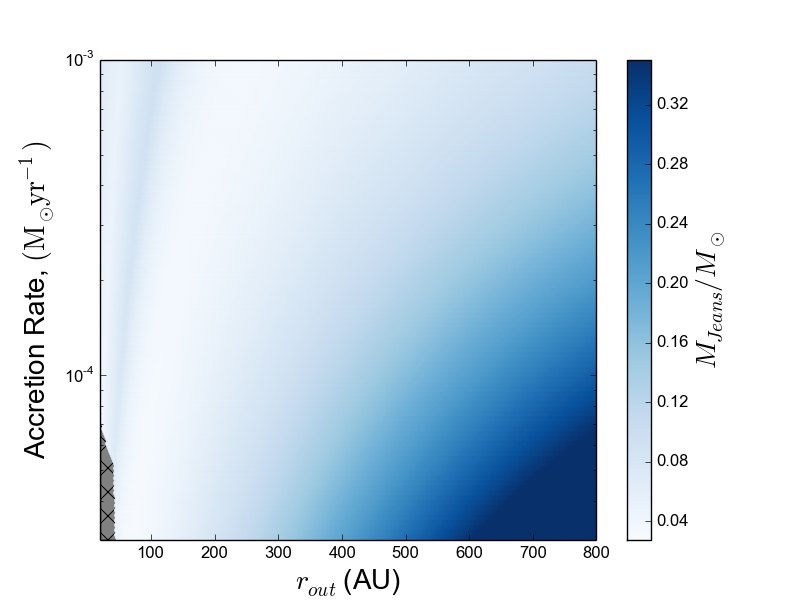} \\
  \end{array}$
  \caption{As Figure \ref{fig:MM1_34msol}, but for the IRAS 16547$-$4247 system.  The disc model (and predicted mass for continuum dust emission) both underestimate the observed mass of $\sim 6 \msol$ significantly.  The observed disc temperatures beyond $\sim$ 100 au are also much higher than suggested by our model, suggesting that this disc is kept gravitationally stable by external irradiation.}
  \label{fig:IRAS16547}
\end{figure*}

If the disc is self-gravitating, then it is unlikely to satisfy this assumption (Figure \ref{fig:IRAS16547}).  For the assumed stellar mass of 20\,$\msol$ (and assumed inner radius of 20 au), our models indicate self-gravitating discs will possess a true mass of 3--5\,$\msol$ depending on accretion rate.  Dust thermal emission (at their assumed dust temperature of 250\,K) will only reveal approximately 15 per cent of this mass, with any other observed emission originating in the envelope and other optically thin components.  If we were to attempt to construct a disc model with around 6\,$\msol$ of observable material, it is likely that the true underlying disc mass would exceed the star mass by a factor of two - an unlikely scenario.

\smallskip

Our model temperatures are slightly lower than measured by \citet{zapata_2015}, who measure molecular lines with excitation temperatures above 500\,K within the disc extent.  The inner regions of our disc model reach approximately 500\,K, but fail to exceed 100\,K beyond approximately 100\,au, i.e. we are unable to reproduce the temperatures consistent with such extensive emission at temperatures above 100K.  It seems that this disc is significantly hotter than predicted from models where the star is the principal irradiation source.  We conclude that external irradiation is permitting this disc to be massive and gravitationally stable, and we are therefore reasonably assured that marginally unstable self-gravitating disc models do not present a good fit to IRAS 16547$-$4247.

\subsection{IRAS 20126$+$4104}

This system (composed of a 12\,$\msol$ star surrounded by a 1.5\,$\msol$ disc) has been intensively modelled to explicitly check for gravitational  instability \citep{chen_2016}.  Rather than relying on a simple dust mass calculation from the continuum, the aforementioned authors fit their data (continuum, \methylcyanide\ and \methanol) to a thin flared Keplerian disc plus envelope model, and find their best fit corresponds to a hot, gravitationally stable disc with $Q>3$ at all radii.

\begin{figure*}
  $\begin{array}{cc}
  \includegraphics[width=\columnwidth]{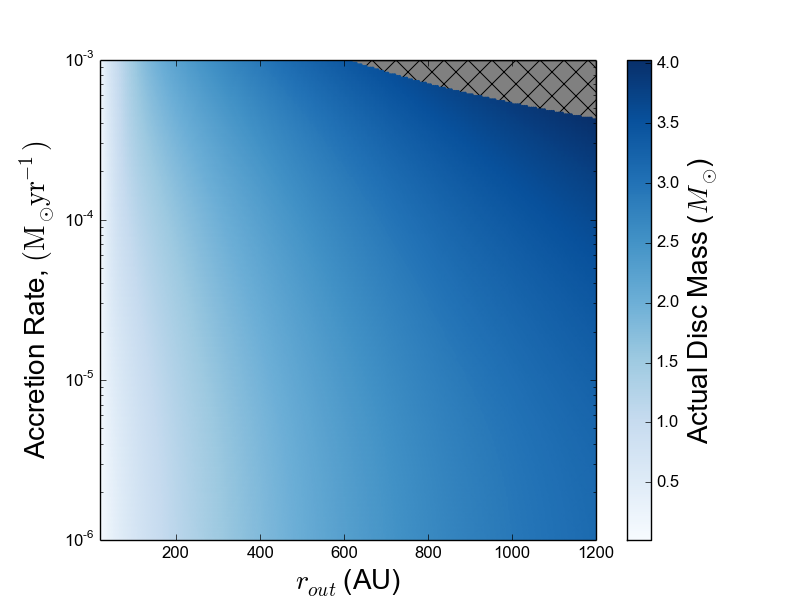} &
  \includegraphics[width=\columnwidth]{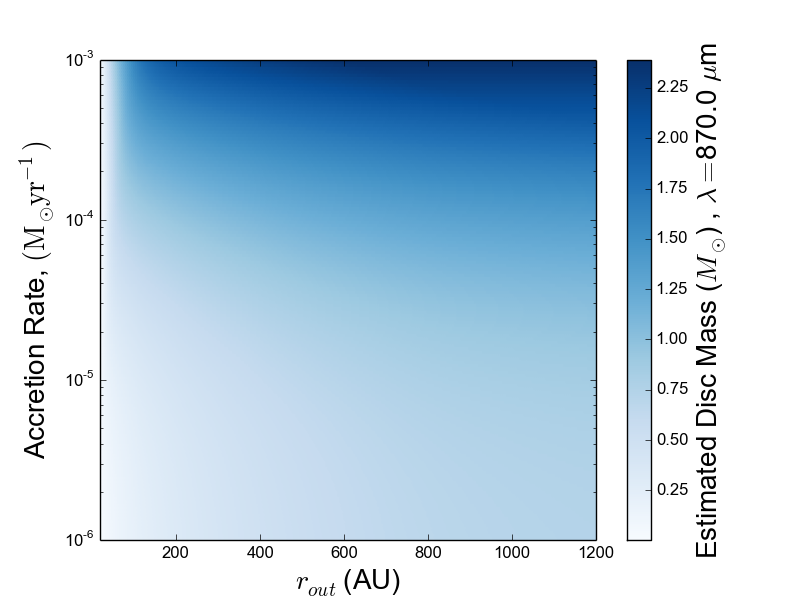} \\
  \includegraphics[width=\columnwidth]{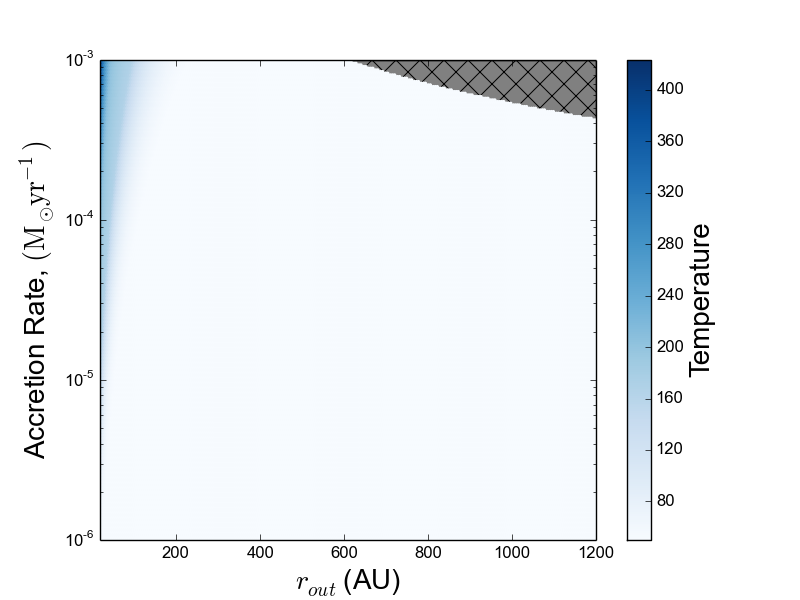} &
  \includegraphics[width=\columnwidth]{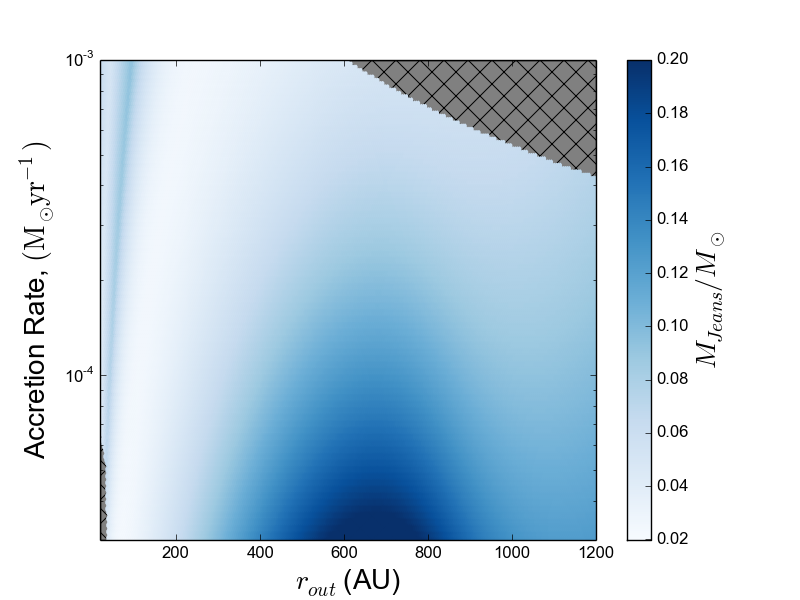} \\
  \end{array}$
  \caption{As Figure \ref{fig:MM1_34msol}, but for the IRAS 20126$+$4104 system.}
  \label{fig:IRAS20126}
\end{figure*}

Our disc models (Figure \ref{fig:IRAS20126}), with an inner radius of 20 au, are broadly consistent with this picture.  Our models fix $Q=2$ at all radii, and as a result produce more massive discs. For their established accretion rate of $3 \times 10^{-5}\, \msol \mathrm{yr}^{-1}$, our models derive a true disc mass of approximately 3\,$\msol$, of which $1.5\,\msol$ is observable in the dust continuum at 870\,$\micron$.  A gravitationally unstable disc is hence a poor description of the data, but this result is of course sensitive to uncertainties in the opacity, and to the accretion luminosity of the star, which we do not explicitly model and is likely to be the cause of our lower disc temperatures (\citet{chen_2016} compute temperatures of approximately 300 K at 200 au, see Figure \ref{fig:IRAS20126}).  \citet{chen_2016} calculate a 27 per cent error in their estimation of $Q$, but it is unclear how they incorporate systematic effects due to opacity uncertainties into this estimate.  Changes to the opacity modify the derived equilibrium disc structure, and hence the derived $Q$.  That being said, we would agree with their statement that this system is just above the marginally unstable regime of $Q<2$.

\section{Conclusions}
\label{sec:conclusions}

We have constructed one dimensional self-gravitating disc models for five disc candidates around massive stars: G11.92$-$0.61 MM1 \citep{Ilee2016}, NGC 6334 I(N) SMA1b \citep{hunter_2014}, AFGL 4176 mm1 \citep{johnston_2015}, IRAS 16547$-$4247 \citep{zapata_2015} and IRAS 20126$+$4104 \citep{chen_2016}.  We show that while in general observed disc masses from continuum dust emission may be quite low, suggesting the systems are not in fact self-gravitating, much of the true disc mass can be masked by the high optical depth of self-gravitating discs even at relatively long (i.e.\ millimetre) wavelengths.

\smallskip

Indeed, we find that our disc models indicate that systems with a star mass of tens of solar masses and an observed disc-to-star mass ratio of $\sim 0.05$ or less can possess true disc-to-star mass ratios of $\sim 0.2$, placing them in the self-gravitating regime, and indeed potentially fragmenting into low mass protostars.

\smallskip

We find two broad categories of disc.  In the first category, the disc is sufficiently massive and cool that marginally stable self-gravitating disc models make accurate predictions for the observed continuum dust emission, while masking a significantly larger true disc mass, potentially prone to fragmentation (G11.92$-$0.61 MM1 and NGC 6334 I(N) SMA1b).  The other three objects belong to the other category, where fragmentation (and gravitational instability generally) is forestalled by sufficient irradiation from the central star and/or external sources.  Our model temperatures at large radii are slightly lower than the observed values for these three sources, which suggests external irradiation or extra sources of luminosity may be important.

\smallskip

Our models exclude the envelope and inner H\,{\sc ii} regions in these systems, and hence our true disc masses will underestimate observed total enclosed masses.  Our intent is not to precisely fit each system, but to illustrate how self-gravitating discs can mask a large fraction of their mass even at millimetre wavelengths.  

\smallskip

If the mass (and hence the stability) of a self-gravitating disc is to be correctly characterised by observations, then one must be cautious when calculating masses using thermal dust emission.  If one assumes the dust is optically thin, then one is at risk of dramatically underestimating the total disc mass.  A fuller analysis of the complete disc structure, ideally with radiative transfer modelling of both the dust and gas emission, is necessary \citep[see e.g.][]{johnston_2015,chen_2016}.

\smallskip

In light of this, we suggest that disc candidates around massive stars that have measured Toomre $Q$ values $<$ 3--5, or measured disc-to-star mass ratios of order $0.05$ or higher, should be modelled carefully to ensure that disc masses have been correctly inferred, and studied for signs of fragmentation.  We note that G11.92$-$0.61 MM1 and NGC 6334 I(N) SMA1b are perhaps the most promising candidates for self-gravitating discs, given the apparent youth of the objects, and the fact a simple self-gravitating disc model agrees well with observations.  As such, we recommend they be studied at higher resolution and sensitivity (such as with ALMA or NOEMA) to search for substructure that is indicative of gravitational instability, and perhaps even low mass star formation via disc fragmentation.

\section*{Acknowledgements}

We would like to thank Katharine Johnston for useful discussions.  DHF gratefully acknowledges support from the ECOGAL project, grant agreement 291227, funded by the European Research Council under ERC-2011-ADG.  JDI gratefully  acknowledges  support  from  the  DISCSIM  project,  grant agreement  341137,  funded  by  the European  Research  Council  under ERC-2013-ADG.  CJC acknowledges support from STFC grant ST/M001296/1.  This  research  has  made  use  of  NASA's  Astrophysics  Data  System Bibliographic  Services.




\bibliographystyle{mnras} 
\bibliography{stability}







\bsp	
\label{lastpage}
\end{document}